\documentclass[twocolumn, aps, rmp, amsmath, amssymb, nofootinbib, superscriptaddress, longbibliography, floatfix, table-of-contents, a4paper, accepted=2020-07-09]{quantumarticle}

\pdfoutput=1
\usepackage[pdftex]{graphicx}
\usepackage{mathrsfs}
\usepackage{url}
\usepackage[colorlinks]{hyperref}
\usepackage{amsmath}    
\usepackage{type1cm}
\usepackage{lettrine}
\usepackage[english]{babel}
\usepackage{microtype}
\usepackage{booktabs}
\usepackage[T1]{fontenc}	
\usepackage{braket}
\usepackage{natbib}
\newcommand{\ketbra}[2]{| #1 \rangle \langle #2 |}
\newcommand{\mc}{\mathcal}
\newcommand{\mr}{\mathrm}
\newcommand{\mbb}{\mathbb}
\DeclareMathOperator{\tr}{Tr}

\begin{document}

\title{A robust W-state encoding for linear quantum optics }

\author{Madhav Krishnan Vijayan }
\email{mkv.215@gmail.com}
\affiliation{Centre for Quantum Software \& Information (UTS:QSI), University of Technology Sydney, Sydney NSW, Australia}
\orcid{0000-0002-7863-7437}

\author{Austin P. Lund}
\email[]{a.lund@uq.edu.au}
\affiliation{Centre for Quantum Computation \& Communications Technology, School of Mathematics \& Physics, The University of Queensland, St Lucia QLD, Australia}
\orcid{0000-0002-1983-3059}

\author{Peter P. Rohde}
\email[]{dr.rohde@gmail.com}
\homepage{http://www.peterrohde.org}
\affiliation{Centre for Quantum Software \& Information (UTS:QSI), University of Technology Sydney, Sydney NSW, Australia}
\orcid{0000-0002-5814-7289}


\frenchspacing

\begin{abstract}
Error-detection and correction are necessary prerequisites for any scalable quantum computing architecture. Given the inevitability of unwanted physical noise in quantum systems and the propensity for errors to spread as computations proceed, computational outcomes can become substantially corrupted. This observation applies regardless of the choice of physical implementation. In the context of photonic quantum information processing, there has recently been much interest in \textit{passive} linear optics quantum computing, which includes boson-sampling, as this model eliminates the highly-challenging requirements for feed-forward via fast, active control. That is, these systems are \textit{passive} by definition. In usual scenarios, error detection and correction techniques are inherently \textit{active}, making them incompatible with this model, arousing suspicion that physical error processes may be an insurmountable obstacle. Here we explore a photonic error-detection technique, based on W-state encoding of photonic qubits, which is entirely passive, based on post-selection, and compatible with these near-term photonic architectures of interest. We show that this W-state redundant encoding techniques enables the suppression of dephasing noise on photonic qubits via simple fan-out style operations, implemented by optical Fourier transform networks, which can be readily realised today. The protocol effectively maps dephasing noise into heralding failures, with zero failure probability in the ideal no-noise limit. We present our scheme in the context of a single photonic qubit passing through a noisy communication or quantum memory channel, which has not been generalised to the more general context of full quantum computation.
\vspace{0.5cm}


\end{abstract}

\maketitle

\tableofcontents

\section{Introduction}

Within quantum information processing systems, the ability to detect errors is an absolute prerequisite for the road towards fault-tolerance. In the standard approach to fault-tolerant quantum computing, one first constructs error-detection circuits, upon which we build error-correction capabilities, finally revisiting the construction to ensure error transversality, facilitating recursive nesting of the protocol to suppress error rates \cite{bib:Shor96FT, bib:Preskill98FT, bib:NielsenChuang00}. In the absence of the initial error detection stage, such a construction for mitigating errors cannot function.

The standard framework when considering quantum error-correction is in the context of universal quantum computation \cite{bib:KLM01}. Given that it is universal, multiple levels of error correcting codes can be implemented. In general this requires large, but sub-exponential, resource overheads each with sub-threshold error rates. Although such constructions are essential for realising the full potential of quantum computing, it remains a distant target. Hence there is currently a pursuit to find utility for achievable near-term devices with post-classical capabilities, even if not universal \cite{Harrow2017, Lund2017}. This has lead to the alternative target where universality is discarded as a requirement and the sole purpose is demonstrating some form of quantum computational advantage with pragmatically reasonable resources. Two examples for such paradigms whose quantum power is proven by links to widely presumed structures in computational complexity theory are IQP (Instantaneous Quantum Polynomial), and boson-sampling devices, both examples of so-called \textit{sampling problems}. IQP is the class of sampling problems consisting of commuting gates acting on qubits that are prepared and measured in superposition basis relative to that of the eigenbasis of the commuting gates \cite{Shepherd2009, Bremner2011}. Boson-sampling is the set of problems that can be constructed from the preparation and measurement of individual bosons subject to evolution via passive linear interferometers \cite{bib:Aaronson11, Aaronson2014}.

In the development of classical hardness arguments for these restricted models, the consideration of errors under a trace-norm induced distance has been of prime importance. Sampling from these distributions with bounded error (which is actually an input to the problem definition) is called \textit{approximate sampling}. The arguments for the classical computational hardness of approximate sampling do not utilize any form of extra resources to deal with errors as the main purpose of these models was the minimisation of quantum resources. Therefore, in standard error analysis for restricted modes the objective has been to find the scaling relationships between trace-norm distance and the parameters defined within the error process (e.g. loss rates, mode overlaps, unitary noise, etc.) \cite{bib:Rohde12_tol,Aaronson2016,Shchesnovich2014,Tichy2015,Leverrier2015,Kalai2014,Arkhipov2015,RahimiKeshari2016}. However, some quantum resources come cheaper than others within these models. In particular, additional modes prepared with vacuum states within the boson-sampling paradigm are considered to have much lower cost than additional modes prepared with single photons. However, given that this model is \textit{passive}, one may suspect that it is not possible to perform any kind of error correction without leaving the constraints of the model, and hence dealing with errors defaults back to the requirements associated with universal models.

Marshman \emph{et al.} \cite{Marshman2018} have shown that, for boson-sampling, it is possible to detect the presence of random phase errors without leaving the paradigm and that the conditional state on detecting the error has a lower error than would otherwise be the case.  This was done using a redundant encoding of the passive linear interferometer with a particular network chosen for encoding and decoding of input single photons.  The presence of the photon within a particular mode was used as the error detection mechanism.  Devices requiring higher photon numbers could be accommodated by parallel combinations of single photon interferometers. This is distinct from the considerations of \cite{Arkhipov2015} for errors within unitary networks as there it was assumed that there was no redundancy utilizing additional resources.  

In this paper, we extend this result by considering single photon encoding that involve W-state path entanglement encoding of photonic qubits encoded in dual-rail form. These states can be generated from single photons through passive linear interferometers, and resemble a generalisation of an optical fan-out operation, having desirable properties for error correction such as the maintenance of path entanglement when single systems are lost. The expansion in mode number can be conceptually related to conventional error-correction schemes based on redundancy, such as Shor's original 3-qubit code \cite{bib:Shor95}. We show that this encoding yields an improvement on local phase shift errors \cite{bib:RohdeRalph06} much like that of the previous work but also show that photon loss is the constraining factor in the heralded fidelity for this localised noise model. We also show that this performance is independent of the type of distribution underlying the random phase errors provided that the errors acting on different modes are independent (i.e., no correlated errors), identical (all modes are treated the same)  and the characteristic function for the distribution is well defined. Under these conditions 
any level of encoding will improve fidelity when conditioned on detecting no error and with a large enough encoding we can fully mitigate the dephasing error.


To present our results we will first discuss different classes of multipartite entangled states and elaborate on why W-states are a good candidate for encoding and define the W-basis in section \ref{sec:multi_ent}. In section \ref{sec:protocol} we introduce our W-state based encoding using only linear optics and single-photon inputs and describe how to post-select to filter our error. In this section we will also describe the linear optics error model that we will consider. Then in section \ref{sec:err_hrld} we compute the success probability of the protocol to succeed and compute the fidelity of the output logical state with the input logical state and show that the performance improves as the level of encoding increases. We discuss how to implement qubit gates on the logical qubit while in the W-state encoding in section \ref{sec:qubit_gates}. Finally and we will discuss some implications of this work and finally make some concluding remarks in section \ref{sec:discussion}.

Importantly, we present an elementary economic argument in Sec.~\ref{sec:economics} for the merit of our scheme from the perspective of engineering economics that, regardless of the state of precision engineering, this approach is likely to be economically justified is some regimes, complimentary to investment into improved precision engineering. This simple observation is based on the intuitive notion that investment into enhanced engineering precision scales exponentially with precision requirements, whereas redundancy scales roughly linearly, in economic overhead, from which the inevitability arises of there being a crossover point between how resources should be allocated to maximise economic efficiency.

\section{Conceptual basis -- Redundancy \& entanglement robustness}\label{sec:multi_ent}

An inherent feature of any kind of multi-qubit entangled state is that, by virtue of its entanglement, loss or decoherence of a single constituent qubit diminishes its degree of entanglement, similarly reducing its purity (or conversely, increasing its collective entropy). Some entangled states are more robust than others in this respect and, as discussed below, the W-states are a quintessential example of entangled states with this robustness property.  Note that the resultant state following a partial trace operation upon a qubit (equivalent to loss when using single photon encoding) is independent of anything done only to traced out qubits prior to the partial tracing operation. Therefore considering loss via partial trace is completely sufficient to understand the worst-case degradation of an entangled state under any kind of local noise process.

\subsection{GHZ states}

The worst-case scenario is the GHZ state, a maximally-entangled $n$-qubit state of the form,
\begin{align}
    \ket{\mathrm{GHZ}_n} = \frac{1}{\sqrt{2}}(\ket{0}^{\otimes n} + \ket{1}^{\otimes n}),
\end{align}
whereby all qubits are collectively perfectly correlated. That is, measurement of any one (in the computational $Z$-basis), reveals the equivalent measurement outcome of all others. However, this directly implies that losing access to this information similarly implies loss of knowledge of the others. Loss or dephasing directly correspond to such loss of information. For this reason, dephasing a single qubit, or losing it outright, implies complete decoherence of the entire $n$-qubit state. Specifically, partial tracing out a single qubit from a GHZ state leaves behind the hopelessly mixed stated,
\begin{align}
    \tr_i(\ket{\mathrm{GHZ}_n}\bra{\mathrm{GHZ}_n}) &= \frac{1}{2}\ket{0}^{\otimes n-1}\bra{0}^{\otimes n-1} \nonumber\\
    &+\frac{1}{2}\ket{1}^{\otimes n-1}\bra{1}^{\otimes n-1},
\end{align}
where the partial trace is performed upon any qubit $i$.

\subsection{Cluster states}

Cluster (or graph) states \cite{bib:Raussendorf01, bib:Raussendorf03} are a highly useful class of states, enabling universal quantum computation using the measurement based model for quantum computing (MBQC). Despite being more computationally useful than GHZ states, they are far less entangled, and hence far more robust against localised noise processes. For example, by measuring out the immediate neighbours of a lost qubit from within a graph state, a reduced, yet perfect graph state is recovered, given by the sub-graph of the original graph, with the neighbourhood of the lost qubit removed.

\subsection{W-states}

An especially robust (and so far not particularly useful) class of entangled states are the W-states \cite{Zeilinger1997a, Dur2000}, given by the equal superposition of a single excitation across $n$-sites. In qubit form this can be expressed,
\begin{align}\label{eq:wn_def}
    \ket{W_n} &= \frac{1}{\sqrt{n}}(\ket{1,0,0,\dots} + \ket{0,1,0,0,\dots} \nonumber \\
    &+ \ket{0,0,1,0,\dots}  + \ket{0,0,0,1,\dots} + \dots).
\end{align}
Alternately, this can be expressed in terms of creation or excitation operators, $\hat{a}^\dag_i$ for the $i$th site,
\begin{align}\label{eq:W_creation}
    \ket{W_n} = \frac{1}{\sqrt{n}} \sum_{i=1}^n \hat{a}^\dag_i \ket\Omega,
\end{align}
where $\ket\Omega$ is the collective ground or optical vacuum state. The latter representation is the one we will focus on here, given its direct applicability to photonic encoding.
 
These states exhibit complete permutational symmetry under qubit interchange. That is, the state is invariant under any permutation $\hat\pi\in S_n$ in the symmetric group,
\begin{align}
    \hat\pi \ket{W_n} = \ket{W_n}.
\end{align}
Tracing out a single qubit from a W-state yields,
\begin{align}
    \tr_i(\ket{W_n}\bra{W_n}) &= \frac{n-1}{n}\ket{W_{n-1}}\bra{W_{n-1}} \nonumber\\
    &+ \frac{1}{n}\ket{0}^{\otimes n-1}\bra{0}^{\otimes n-1}.
\end{align}
That is, upon loss of a single qubit, with probability \mbox{$p=(n-1)/n$} it simply undergoes a reduction in its level of encoding to a $\ket{W_{n-1}}$ state, preserving its W-type structure entirely, otherwise collapsing to the $\ket{0}^{\otimes n-1}$ state. This implies that that for large $n$, W-states are extremely robust (indeed almost invariant) against single-qubit loss. As discussed earlier, this directly implies similar single-qubit robustness against other noise channels.

Note that atomic ensemble qubits \cite{bib:DuanLukin01} are a direct alternate physical manifestation of W-type encoding, whereby an ensemble (or cloud) of collectively-addressed atomic qubits undergo \textit{collective excitation}, mathematically of the form given in Eq.~\eqref{eq:W_creation}. This approach to realising physical qubits has attracted much attention, especially as good candidates for quantum memories, given their notably high coherence lifetimes, often at room temperature, which can be intuitively associated with the described robustness of their underlying W-type entanglement structure -- if a few atoms go missing from the cloud, little is lost.

The $n$-qubit W-state can be easily generalised to an entire orthonormal W-basis, by appropriately manipulating the phase relationships within the $n$ terms in the superposition. One way in which to choose these phases is by taking the elements from the Quantum Fourier Transform (QFT) matrix, or generalised Hadamard matrices, both of which have equal $1/\sqrt{n}$ amplitudes across all matrix entries, with phase relationships ensuring orthonormality.

These different phase relationships do not change the earlier observations about the states' robustness against local noise. This immediately leads to the intuition, that by choosing such a W-basis for encoding logical qubits, the encoded logical qubit must inherit via linearity these same robustness characteristics. This makes them a direct candidate for optical encoding, given that photonic implementation of QFT mappings may be implemented via passive linear optics, in the absence of any active control, and is realisable with today's technology integrated wave-guide architectures across a large number of modes.

\section{Protocol}\label{sec:protocol}

\begin{figure}[!htbp]
    \includegraphics[width=\columnwidth]{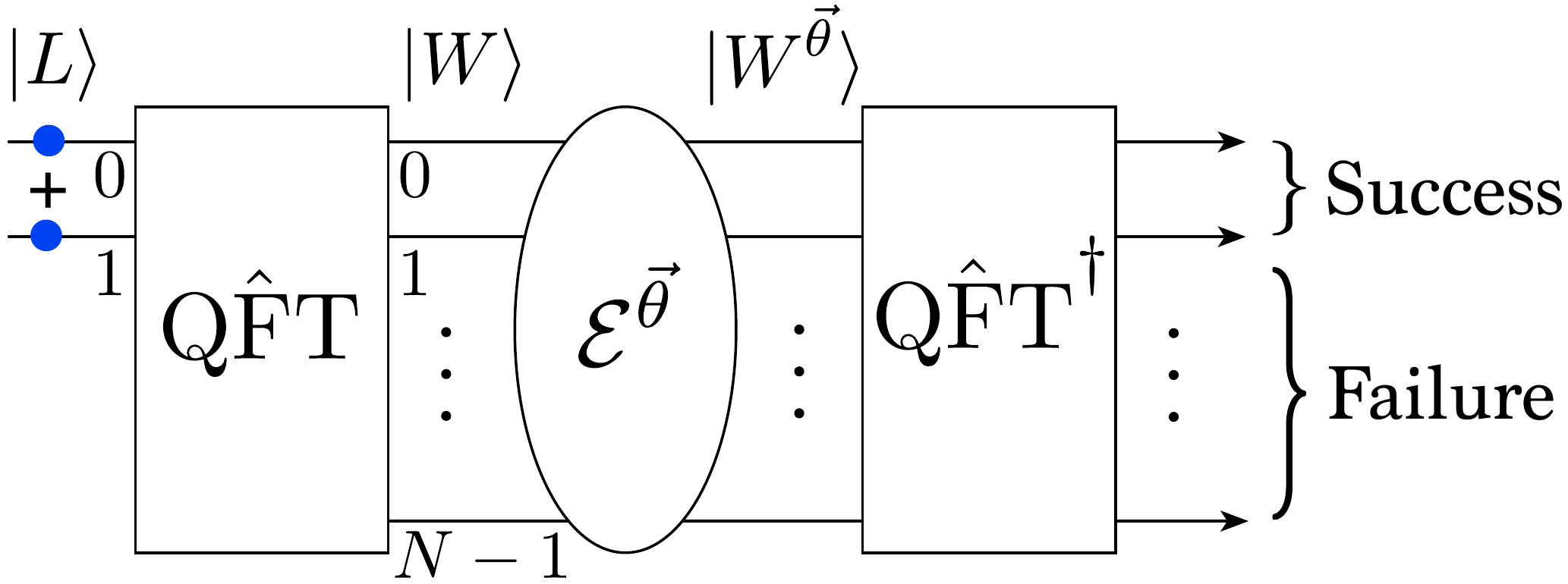}
    \caption{Photonic W-state error-correction and -detection protocol. Encoding of a single dual-rail photonic qubit proceeds via a Quantum Fourier Transform ($\hat{QFT}$), which maps the 2-mode encoding across a larger number of redundant modes. The independent dephasing noise channel is denoted by $\mathcal{E}$. Decoding proceeds via the inverse Quantum Fourier Transform ($\hat{QFT}^\dag$). Post-selection upon detecting the single photon within the desired 2 output modes defining the single qubit, projects the logical state into one with reduced noise action.}
    \label{fig:protocol}
\end{figure}

The error-detection and correction protocol is shown in Fig.~\ref{fig:protocol}. Consider $N$ optical modes, the first two of which contain a single photon state, defining a dual-rail-encoded photonic qubit. This qubit can defines a \textit{logical} qubit,
\begin{align}
    \ket{L} &= \alpha\ket{0}_L + \beta\ket{1}_L \nonumber \\
    &= (\alpha \hat{a}^\dag_0 + \beta \hat{a}^\dag_1)\ket{\Omega},
\end{align}
where $\ket{\Omega}$ is the $N$ mode vacuum state. To W-encode the logical qubit we pass the $N$ modes through a linear optical network implementing the $N$-mode quantum Fourier transform,
\begin{align}\label{Eq:Wi_op_def}
    \hat{a}_i^\dag \to \hat{W}^\dag_i = \sum_{j=0}^{N-1}  \hat{\mathrm{Q}}_{ij} \hat{a}^\dag_j,
\end{align}
where,
\begin{align}
    \hat{\mathrm{Q}}_{jk} = \dfrac{\omega_N^{jk}}{\sqrt{N}}; \hspace{1cm} j, k \in \{ 0,1,\dots,N-1 \},
\end{align}
are matrix elements of the $N$-dimensional quantum Fourier transformation operator $\hat{\rm{Q}}$ with $\omega_N = e^{\frac{2\pi i}{N}}$. 
This transforms the \textit{logical} qubit to the \textit{encoded} qubit,
\begin{align}
\ket{W} = (\alpha \hat{W}^\dag_0 + \beta \hat{W}^\dag_1)\ket{\Omega},	
\end{align}
which represents the same state of quantum information, but in expanded form. Next the W-encoded state passes through a noisy channel that independently adds random phases to each optical mode,
\begin{align}
	\hat{a}^\dag_j \to e^{i\theta_j} \hat{a}^\dag_j,
\end{align}
where the $\theta_j$ represent random variables, whose distribution is considered arbitrary at this point, that form a vector $\vec{\theta}$ describing the phases applied to each mode.
The state $\ket{W^{\vec\theta}}$ denotes the W-encoded state following application of the phase noise channel. We now apply the decoding operation (the inverse QFT operation), and the first two output modes represent the decoded logical state, $\hat\rho_L$. Because of the noise in the channel we are not guaranteed to observe the photon strictly within the first two modes. Thus we post-select and treat cases where photons are found in the other modes as heralding a failure. The intuition is then that for the heralded success cases the phase noise errors would have been filtered out.

The fidelity of the decoded state compared to the input logical qubit $\ket{L}$ is
\begin{align}
F_N = \bra{L}\hat\rho_L\ket{L},	
\end{align} 
where $\ket{L}$ is implicitly a function of the superposition parameters $\alpha$ and $\beta$. 
Note that the overlap between two states is invariant under common unitary operations. As the encoding and decoding operations are unitary it suffices to consider the fidelity of the W-encoded state,
\begin{align}\label{eq:Fid_particular}
F_N &= \braket{ W | W^{\vec\theta}} \braket{ W^{\vec\theta} | W},
\end{align}
where $F_N$ is used here to show that the fidelity will depend on the number of modes used for the encoding $N$.
Eq.~\eqref{eq:Fid_particular} assumes knowledge of the phase errors in each mode as represented by $\vec\theta$ but these are of course unknown. However we can model them as independent random variables acting on each mode separately according to some arbitrary distribution $p$,
\begin{align}
\label{arb-dist}
	p(\vec\theta) = \prod_{j=0}^{N-1} p_i(\theta_j),
\end{align}
where $p_j$ is the distribution for mode $j$.
The encoded state after application of the error channel on average is given by,
\begin{align}
\hat\rho_W = \int p(\vec\theta) \ket{W^{\vec\theta}}\bra{W^{\vec\theta}}\, d\vec\theta.
\end{align}
The fidelity between the output and input of the error channel is given by,
\begin{align}
	F_N 
    &= \bra{W} \left( \int p(\vec\theta)  \ketbra{W^{\vec\theta}}{W^{\vec\theta}} d\vec\theta \right) \ket{W}. \label{eq:Fid_integral}
\end{align}

As with all quantum operations, the noise channel is a linear map on the state space. Let the channel map be denoted by $\mc{L}_{\vec\theta}$, then we have for the encoded qubit state,
\begin{align}
\ket{W^{\vec\theta}} &= \mc{L}_{\vec\theta}(\ket{W}) = \alpha \mc{L}_{\vec\theta}(\ket{W_0}) + \beta \mc{L}_{\vec\theta}(\ket{W_1})\nonumber \\
&= \alpha \ket{W_0^{\vec\theta}} + \beta \ket{W_1^{\vec\theta}},
\end{align}
where $\ket{W_k} = \hat{W}_k^{\dagger} \ket{\Omega}$, following the definition in Eq.~\eqref{Eq:Wi_op_def},
\begin{align}\label{Eq:Wi_state_def}
 \ket{W_k^{\vec{\theta}}} =  \sum_{j}^{N-1} \hat{\mr{Q}}_{kj}  e^{i\theta_j}a^{\dagger}_j \ket{\Omega}. 
 \end{align}
 
These equations can now be used to compute $\hat\rho_W$,
\begin{align}
\hat\rho_W &= \int p(\vec\theta) \left( \ketbra{W^{\vec\theta}}{W^{\vec\theta}} \right) d\vec\theta \nonumber \\
&= \int p(\vec\theta) \left(\sum\limits_{i,j =0}^1 c_{i}c^*_{j}\ketbra{W_i^{\vec\theta}}{W_j^{\vec\theta}} \right) d\vec\theta,
\end{align}
where $c_{0} = \alpha$ and $c_{1} = \beta$. Substituting the definition of $\ket{W_i^{\vec\theta}}$ from Eq.~\eqref{Eq:Wi_state_def} and defining,
\begin{align}
    k_{m,n} = (\alpha + \beta\omega_N^m)(\alpha + \beta \omega_N^n)^*,
\end{align}
we obtain,
\begin{widetext}
\begin{align}
    \hat\rho_W = \frac{1}{N}
   \int p(\vec\theta) \begin{pmatrix}
    k_{0,0} &k_{01}e^{i(\theta_0-\theta_1)}  &k_{0,2}e^{i(\theta_0-\theta_2)} &\dots &k_{0,(N-1)}e^{i(\theta_0-\theta_{N-1})} \\
    k_{1,0}e^{i(\theta_1 - \theta_0) } &k_{1,1} &k_{12}e^{i(\theta_1 - \theta_2)} &\dots \\
    k_{2,0}e^{i(\theta_2 - \theta_0)} &k_{2,1}e^{i(\theta_2 - \theta_1)} &k_{2,2} &\dots \\
    \vdots & & &\ddots \\
    k_{(N-1),0}e^{i(\theta_{N-1} - \theta_0)} & & & &k_{(N-1),(N-1)}
    \end{pmatrix}d\vec\theta,
\end{align}
\end{widetext}
where the matrix within the integral represents the state after the noise channel in the photon number basis.
The characteristic function of a probability distribution $p(x)$ is defined as,
\begin{align}
\phi_{p(x)}(z) = \int\limits_{-\infty}^{\infty} p(x)e^{ix z} dx.
\end{align}
If we assume all $\theta_j$ are identically and independently distributed as $p(\theta)$ then we have,
\begin{align}\label{eq:char_func}
    \lambda = \int p(\vec\theta) e^{i(\theta_j - \theta_k)} d\vec\theta = |\phi_{p(\theta)}(1)|^2,
\end{align}
whenever the indices $j$ and $k$ are different.
Thus,
\begin{align}
    \hat\rho_W &= \frac{\lambda}{N}\sum_{i, j = 0}^{N-1} (1 - \delta_{ij}) k_{i,j} \ketbra{i}{j}  + \frac{1}{N}\sum_{i}^{N-1} k_{i,i} \ketbra{i}{i} \nonumber \\
    &= \lambda\left(\ketbra{W}{W} - \Delta(\ketbra{W}{W}) \right) + \Delta(\ketbra{W}{W}) \nonumber \\
    &= \lambda\ketbra{W}{W} + (1 - \lambda)\Delta(\ketbra{W}{W}), \label{Eq:depahsing_characteristic}
\end{align}
where we have used the fact that,
\begin{align}
    \sum\limits_{i,j = 0}^{N-1} \frac{k_{i,j}}{N}\ketbra{i}{j} = \ketbra{W}{W},
\end{align}
and $\Delta$ is the completely dephasing map in the photon number basis defined as,
\begin{align}
    \Delta(\hat\rho) = \sum_i \braket{i | \hat\rho | i}\ketbra{i}{i}.
\end{align}
We can see from Eq.~\eqref{Eq:depahsing_characteristic} that the our error channel is essentially a dephasing channel with dephasing parameter $\lambda$.
To analyze our protocol further we will  choose a particular error model by assuming that the phase error in each mode is distributed as a Gaussian with mean $\mu$ and variance $\delta^2$\footnote{This assumption allows for values of $\theta$ that are larger than single multiples of $2\pi$, but the theory used here does not need to be changed to incorporate this.  The operations used in defining the phase shift channel are periodic and hence having larger value of phase does not invalidate this description.  However, it does mean that there is no unique probability density function for any given distribution on the range $[0,2\pi)$.},
\begin{align}
p(\theta) = \frac{1}{\sqrt{2\pi}\delta}e^{-\frac{(\theta - \mu)^2}{2\delta^2}}. \label{Eq:guassian_err}
\end{align}
This is a natural choice when we do not have any knowledge about the nature of the processes that generates the errors beyond that many underlying random distributions average to give a final contribution to the error (\`a la the central limit theorem). The characteristic function of a normal distribution is given by,
\begin{align}
    \phi_{p(\theta)}(z) = e^{-\frac{\delta^2 z^2}{2} +i\mu z}.
\end{align}
This leads to $\lambda = e^{-\delta^2}$ and,
\begin{align}
\hat\rho_W = e^{-\delta^2} \ketbra{W}{W} + (1 - e^{-\delta^2})\Delta\left( \ketbra{W}{W} \right).
\label{Eq:Rho_W_result}
\end{align}
We can interpret the error channel as performing the identity with probability $e^{-\delta^2}$ and applying the Fock basis dephasing operator with probability $(1 - e^{-\delta^2})$. This channel is thus a dephasing channel with probability of no error occurring $p = e^{-\delta^2}$. In practical terms, the variance $\delta^2$ of the phase error will depend on the physical implementation of the quantum channel. For fibre-optic cables we would generally expect the variance to increase with the length of the cable $L$ or equivalently the propagation time of the photon in cable $t_p = L/v$, where $v$ is the propagation velocity of the photon in the fibre. If we model the variance as increasing linearly with propagation time, i.e,
\begin{align}\label{eq:delta_t2}
\delta^2 = \frac{t_p}{T_2},
\end{align}
where $T_2$ is a constant defining a characteristic time for the dephasing channel, we can write down our error channel in the standard dephasing channel notation as,
\begin{align}
\hat{\rho}_W &= \mathcal{E}^{\text{dephasing}}_{t_p}(\ketbra{W}{W})\nonumber\\
&= e^{-t_p / T_2 } \ketbra{W}{W} + (1 - e^{-t_p / T_2 })\Delta(\ketbra{W}{W}).
\end{align}

\section{Error heralding}\label{sec:err_hrld}

In implementing the protocol as described in Fig.~\ref{fig:protocol}, we can perform the post-selection in two different ways:

\begin{itemize}
\item \textbf{Presence heralding}: Success is assumed based upon the detection of exactly one photon between the output modes 0 and 1, which define the logical qubit space. Note that in the absence of quantum non-demolition measurements, this is necessarily destructive, limiting its applicability. The heralding operator is effectively the projector
\begin{align}
    \hat\Pi_\mathrm{presence} = \hat{a}_0^\dag\ket\Omega\bra\Omega\hat{a}_0 + \hat{a}_1^\dag\ket\Omega\bra\Omega\hat{a}_1.
\end{align}
\item \textbf{Absence heralding}: Success is inferred via the detection of no photons in any of the remaining modes outside the logical qubit space. This is non-destructive on the logical qubit, broadening its utility. However, photon loss contributes to the occurrence of this signature, implying higher error rates on the remaining logical qubit. The heralding operator is equivalently a projection given by
\begin{align}
    \hat\Pi_\mathrm{absence} = \mbb{\hat{I}} - \sum_{i=2}^{N-1} \hat{a}_i^\dag\ket\Omega\bra\Omega\hat{a}_i.
\end{align}
\end{itemize}

\subsection{Heralding probability}

\subsubsection{Absence heralding}

We define the absence heralded probability $P_{H_a}$ as the probability that no photons are detected in modes 2 to ($N-1$) and $\hat\rho_\mathrm{out}$ to be the $N$-mode state of the system at the end of the protocol before the final measurement. Assuming a uniform loss model parameterized by $\eta$, for our choice of input states the probability of detecting the photon in mode $m$ is
\begin{align}
    (1 -\eta) \cdot \braket{\Omega|\hat{a}_m \hat\rho_\mathrm{out} \hat{a}_m^{\dagger}|\Omega},
\end{align}
where $\ket{\Omega}$ is the global vacuum state. Using this expression for the probability of detection under loss we can see that,
\begin{align}
 P_{H_a} &= \text{Pr}(\text{Photon is in modes 0 or 1}) \nonumber\\
 &+ \text{Pr}(\text{Photon is in modes 2-($N-1$)}) \nonumber\\
 &\times \text{Pr}(\text{Loss in modes 2-($N-1$)}), \nonumber \\
 &= \sum_{i=0}^1\braket{\Omega |\hat{a}_i \hat\rho_\mathrm{out} \hat{a}_i^{\dagger}| \Omega} + \eta \left(1 - \sum_{i=0}^1\braket{\Omega |\hat{a}_i \hat\rho_\mathrm{out} \hat{a}_i^{\dagger}| \Omega} \right) \nonumber,\\
&= \eta + (1 -\eta)  \left( \sum_{i=0}^1 \braket{\Omega |\hat{a}_i \hat\rho_\mathrm{out} \hat{a}_i^{\dagger}| \Omega} 
 \right) \nonumber ,\\
&= \eta + (1-\eta) \left( \sum_{i=0}^1 \braket{\Omega |\hat{a}_i {\rm\hat{Q}}^{\dagger}\hat\rho_{W}{\rm\hat{Q}} \hat{a}_i^{\dagger}| \Omega} \right)\nonumber ,\\
&= \eta + (1 -\eta) \left( \sum_{i=0}^1 \braket{W_i | \hat\rho_{W} | W_i} \right)  . \label{Eq:herald_prob}
 \end{align}
Using Eq.~\eqref{Eq:depahsing_characteristic} we can write,
\begin{align}\label{eq:rho_W_braket}
\braket{W_k |\hat\rho_W | W_k} &= \lambda \left| \braket{W_k | W} \right|^2 \nonumber\\
&+ (1 - \lambda)\braket{W_k | \Delta(\ketbra{W}{W})| W_k}.
\end{align}
The first term on the R.H.S. of Eq.~\eqref{eq:rho_W_braket} is simply,
\begin{align}
    \left| \braket{W_0 | W} \right|^2 &= |\alpha|^2,\nonumber\\
    \left| \braket{W_1 | W} \right|^2 &= |\beta |^2,
\end{align}
which are the values of $k$ that preserve the encoded qubit. The second term can be calculated as,
\begin{align}
&\braket{W_k | \Delta(\ketbra{W}{W})| W_k} =
\nonumber \\ & \frac{1}{N^2} \sum_{m, n = 0}^{1}\sum_{j,q,p = 0}^{N-1} \omega_{N}^{(m-n)j + (q - p)k} c_m c_n^* \delta_{pj}\delta_{jq} \nonumber , \nonumber \\
&= \frac{1}{N^2} \sum_{m, n = 0}^{1}\sum_{j = 0}^{N-1} \omega_{N}^{(m-n)j } c_m c_n^* \nonumber \\
&= \frac{1}{N} \sum_{m, n = 0}^{1}\delta_{mn} c_m c_n^ *  = \frac{1}{N},  
\end{align}
where in the first equality we have used the fact that,
\begin{align}
 \ket{W_k} = \sum_{q = 0}^{N-1} 
 \hat{\rm{Q}}_{kq} \hat{a}_q^{\dagger}\ket{\Omega},
\end{align}
and
\begin{align}
\Delta(\ketbra{W}{W}) = \sum\limits_{m, n = 0}^{1} c_mc_n^* \sum_{j = 0}^{N-1} \omega_{N}^{(m - n)j} \hat{a}_j^{\dagger} \ketbra{\Omega}{\Omega} \hat{a}_j .
\end{align}
This implies that the photon in the error state is equally spread over all the modes after decoding. If the state contains an error, the heralding will detect it with a probability of $\frac{N-2}{N}$ and miss it with probability $\frac{2}{N}$. So there will be a linear advantage in error detection with the number of modes.
Substituting these results in Eq.~\eqref{Eq:herald_prob} we get,
\begin{align}
P_{H_a} &= \eta + (1  -\eta) \left[ \lambda|\alpha|^2 + \frac{1}{N}(1 - \lambda) \right] \nonumber \\
&+ (1 - \eta) \left[ \lambda|\beta|^2 + \frac{1}{N}(1 - \lambda)  \right] \nonumber \\
&= \eta + (1 -\eta) \left[ \lambda + \frac{2}{N}(1 - \lambda) \right]. \label{Eq:W_i_rw_wi}
\end{align}
If we assume the Gaussian error model in Eq.~\eqref{Eq:guassian_err} this will reduce to,
\begin{align}
   P_{H_a} =  \eta + (1 -\eta) \left[ e^{-\delta^2} + \frac{2}{N}(1 - e^{-\delta^2}) \right]. 
\end{align}
As the number of modes $N$ increases the heralded probability will decrease, this is because the probability of the error state being in the modes 1 and 2 is inversely proportional to $N$. As we connected the phase error variance to a $T_2$ time via Eq.~\eqref{eq:delta_t2}, we can also reparameterize the loss probability as $\eta = 1 -  e^{-tp/T_1}$. In terms of the $T_1$ and $T_2$ parameters and propagation time $t_p$, the absenence heralded probability can be written as,
\begin{align}
  P_{H_a} &= (1 - e^{-t_p /T_1}) \nonumber\\
  &+ e^{-t_p / T_1} \left[ e^{-t_p / T_2} + \frac{2}{N}(1 - e^{-t_p / T_2}) \right].
\end{align}

\subsubsection{Presence heralding}

The presence heralded case is the case where we post-select on there being no photon loss. The presence heralded fidelity is the probability of getting a photon in modes 1 and 2 and this can be easily seen to be,
\begin{align}
    P_{H_p} = P_{H_a} - \eta.
\end{align}

\subsection{Heralded fidelity}

\subsubsection{Absence heralding}

The absence heralded state is the state in the output modes 0 and 1 when no photons are detected in the modes modes 2 -- $(N-1)$. This can happen in two mutually exclusive ways; either the photon is lost and there is no photon in any mode or there is no loss and our negative measurement of modes 2 -- $(N-1)$ projects the quantum state into the subspace spanned by $a_0^{\dagger}\ket{\Omega}$ and $a_1^{\dagger}\ket{\Omega}$. So, the absence heralded state is given by,
\begin{align}
\hat\rho_{H_a} = (1 - \eta) \frac{\hat\Pi_{0,1} \hat\rho_\mathrm{out} \hat\Pi_{0,1}}{\tr(\hat\Pi_{0,1} \hat\rho_\mathrm{out})} + \eta\ketbra{\Omega}{\Omega},
\end{align}
where,
\begin{align}
    \hat\Pi_{0,1} = a_0^{\dagger}\ketbra{\Omega}{\Omega} a_0 + a_1^{\dagger}\ketbra{\Omega}{\Omega} a_1,
\end{align}
is the projector on to the subspace of modes 1 and 2. But we observe that,
\begin{align}
\tr(\hat\Pi_{0,1} \hat\rho_\mathrm{out}) &= \braket{\Omega | \hat{a}_0 \hat\rho_\mathrm{out} \hat{a}_0^{\dagger} | \Omega} + \braket{\Omega | \hat{a}_1 \hat\rho_\mathrm{out} \hat{a}_1^{\dagger} | \Omega} \nonumber \\
&= \braket{\Omega | \hat{a}_0 \hat{\rm{Q}}^{\dagger}\hat{\rm{Q}} \hat\rho_\mathrm{out} \hat{\rm{Q}}^{\dagger}\hat{\rm{Q}} \hat{a}_0^{\dagger} | \Omega} \nonumber\\
&+ \braket{\Omega | \hat{a}_1 \hat{\rm{Q}}^{\dagger}\hat{\rm{Q}} \hat\rho_\mathrm{out} \hat{\rm{Q}}^{\dagger}\hat{\rm{Q}} \hat{a}_1^{\dagger} | \Omega} \nonumber \\
&= \braket{W_0 |  \hat\rho_{W}  | W_0} + \braket{W_1 | \hat\rho_{W}  |W_1} \nonumber \\
&= \lambda + \frac{2}{N}(1 - \lambda).
\end{align}
The fidelity of the heralded state with our logical input state is given by,
\begin{align}
F_{H_a} &=  \braket{L | \hat\rho_{H_a} | L} \nonumber  \\
&= \braket{L |{\rm\hat{Q}}^{\dagger}{\rm\hat{Q}} \hat\rho_{H_a}  {\rm\hat{Q}}^{\dagger}{\rm\hat{Q}}| L} \nonumber \\
&= (1 - \eta)\frac{\braket{W | \hat\rho_{W} | W}}{\tr(\hat\Pi_{W_0, W_1}\rho_{W})} \nonumber \\
&= \frac{1 - \eta}{\lambda + \frac{2}{N}(1 - \lambda)} \cdot F(\ket{W}, \hat\rho_W),\label{Eq:Fid_compare}
\end{align}
where,
\begin{align}
    \hat\Pi_{W_0,W_1} = \ketbra{W_0}{W_0} + \ketbra{W_1}{W_1},
\end{align}
and in the third equality we have used the fact that $\hat\Pi_{0,1} \ketbra{\Omega}{\Omega} \hat\Pi_{0,1} =0$. We can interpret Eq.~\eqref{Eq:Fid_compare} as saying that heralding improves the output fidelity of our protocol by a factor of,
\begin{align}
    \frac{1 - \eta}{\lambda + \frac{2}{N}(1 - \lambda)}.
\end{align}
We have,
\begin{align}
F(\ket{W}, \hat\rho_W) = \lambda + (1 - \lambda)\braket{W |\Delta(\ketbra{W}{W}) |W}.
\end{align}
Notice that,
\begin{align}
  \braket{W |\Delta(\ketbra{W}{W}) |W} &= \tr(
 \ketbra{W}{W}\Delta(\ketbra{W}{W})) \nonumber \\
 &= \sum_{i = 0}^{N-1} ([\ketbra{W}{W}]_{ii})^2,
  \end{align}
  where, $[\ketbra{W}{W}]_{ii}$ are diagonal elements of the the state $\ketbra{W}{W}$ in the computational basis. We know that,

  \begin{align}
  &[\ketbra{W}{W}]_{ii} = \nonumber\\ 
  & \braket{ \Omega|\hat{a}_{i} \left( \sum_{m,n = 0}^{1} c_mc_n^* \sum_{k,j = 0}^{N-1} \hat{\rm{Q}}_{mk}\hat{a}_k^{\dagger}\ketbra{\Omega}{\Omega}\hat{a}_j \hat{\rm{Q}}^*_{nj} \right) \hat{a}_i^{\dagger}| \Omega } \nonumber \\
&= \frac{1}{N}\sum_{m,n = 0}^{1}\sum_{k, j = 0}^{N-1} c_m c_n^* \omega_{N}^{mk - nj}\delta_{ik}\delta_{ji} \nonumber \\
&= \frac{1}{N}\sum_{m,n = 0}^{1} c_mc_n^*\omega_{N
}^{(m-n)i}. 
  \end{align}

 Using the above expression we obtain,
 \begin{align}\label{eq:summation_calc}
 &\braket{W |\Delta(\ketbra{W}{W}) |W} \nonumber \\
 &= \sum_{i = 0}^{N-1}\left( \frac{1}{N}\sum_{m,n = 0}^{1} c_mc_n^*\omega_{N}^{(m-n)i} \right) \left( \frac{1}{N}\sum_{p,q = 0}^{1} c_pc_q^*\omega_{N}^{(p-q)i} \right) \nonumber \\
 &= \frac{1}{N^2} \sum_{i = 0}^{1} \sum_{\substack{m,n,\\  p, q } = 0}^{1}c_mc_pc_n^*c_q^*\omega_{N}^{(m - n + p - q    )i}  \nonumber \\
&= \frac{1}{N} \sum_{\substack{m,n,\\  p, q } = 0}^{1}c_mc_pc_n^*c_q^* \delta_{m-n+p, q}  \nonumber \\
&= \frac{1}{N} \sum_{\substack{m,n,p  = 0 \\ m - n + p \geq 0} }^1 c_mc_pc_n^*c_{m-n+p}^* \nonumber \\
&= \frac{|\alpha|^4 + |\beta|^4 + 4 |\alpha|^2|\beta|^2}{N} \nonumber \\
&= \frac{1 + 2|\alpha|^2|\beta|^2}{N}.
\end{align}
Therefore, 
\begin{align}
F_{H_a}  = (1 - \eta) \cdot \frac{\lambda + (1 - \lambda)\frac{1 + 2|\alpha|^2|\beta|^2}{N} }{\lambda + \frac{2}{N}(1 - \lambda)}.
\end{align}
In terms of Bloch variables $\theta$ and $\phi$ where $\alpha = \cos\frac{\theta}{2}$, $\beta = e^{i\phi}\sin\frac{\theta}{2}$ as, it can be seen that, 
\begin{align}
F_{H_a} = (1 - \eta) \cdot \frac{e^{-\delta^2} + (1 - e^{-\delta^2})(\frac{2 + \sin^2\theta}{2N} )}{e^{-\delta^2} + \frac{2}{N}(1 - e^{-\delta^2})}.
\end{align}
In the limit of large $N$ the heralded fidelity will approach $(1 - \eta)$. This implies that the only error will be from photon loss. In terms of the dephasing and amplitude damping channel parameters $T_2$ and $T_1$ and a propagation time $t_p$ this can be written as,
\begin{align}
F_{H_a} = e^{-t_p / T_1} \cdot \frac{e^{-t_p/T_2} + (1 - e^{-t_p/T_2})(\frac{2 + \sin^2\theta}{2N} )}{e^{-t_p/T_2} + \frac{2}{N}(1 - e^{-t_p/T_2}) }.
\end{align}

\subsubsection{Presence heralding}

In the presence heralded case, we are post selecting the case where there is no photon loss in the system, so the post measurement state in this scenario will be,
\begin{align}
\hat\rho_{H_p} =  \frac{\hat\Pi_{0,1} \hat\rho_\mathrm{out} \hat\Pi_{0,1}}{\tr(\hat\Pi_{0,1} \hat\rho_\mathrm{out})}
\end{align}
this just improves the fidelity by a factor of $(1 -\eta)$ giving, the fidelity as,
\begin{align}
    F_{H_{p}} = \frac{F_{H_a}}{1 - \eta}.
\end{align}
The heralded probability is plotted in Fig.~\ref{fig:results} as a function of $\delta$ and $T_2$ with fixed values of $\eta$ and $T_1$ respectively. From these plots it is evident that the heralded fidelity improves with the number of modes $N$. The choice of the parameters values $T_1$ and $\eta$ do not influence the ordering of these plots.

\begin{table*}[!htbp]
    \centering
    \begin{tabular}{|c|c|c|}
    \hline
    \textbf{Heralding type} & \textbf{Probability} & \textbf{Fidelity} \\ \hline \hline
Presence    &  $P_{H_p} = (1 - \eta)[\lambda + \frac{2}{N}(1 - \lambda)]$   & $F_{H_p} = \frac{N\lambda + (1 - \lambda)(1 + 2|\alpha\beta|^2)}{(N-2)\lambda + 2}$ \\ 
\hline
Presence ($N\to\infty$)& $P_{H_p} = (1 - \eta)\lambda$ & $F_{H_p}=1$ \\
\hline
Absence & $\eta + P_{H_p}$ & $(1 - \eta) F_{H_p}$ \\
\hline
    \end{tabular}
    \caption{Heralding probabilities and post-selected logical qubit fidelities of a single photon qubit under the W-state encoding protocol, according to the two different modes of post-selection operation. Note that here $\lambda = |\phi_{p(\theta)}(1)|^2$ as defined in equation \eqref{eq:char_func}. 
    }
    \label{tab:analytic_summary}
\end{table*}

\subsection{Probability and fidelity plots}

\begin{figure*}[!htpb]
    \includegraphics[width= \columnwidth]{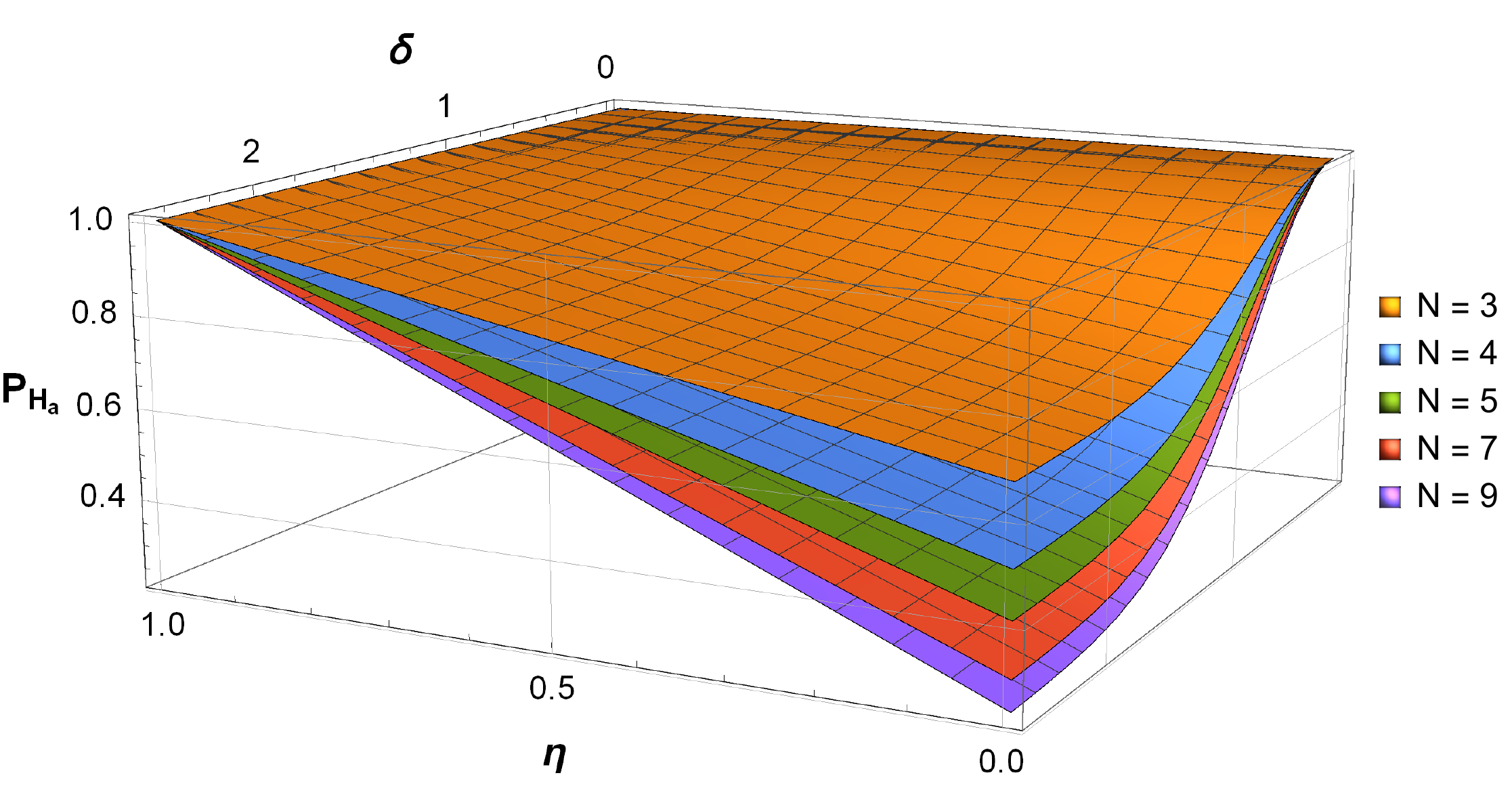}     
    \includegraphics[width= \columnwidth]{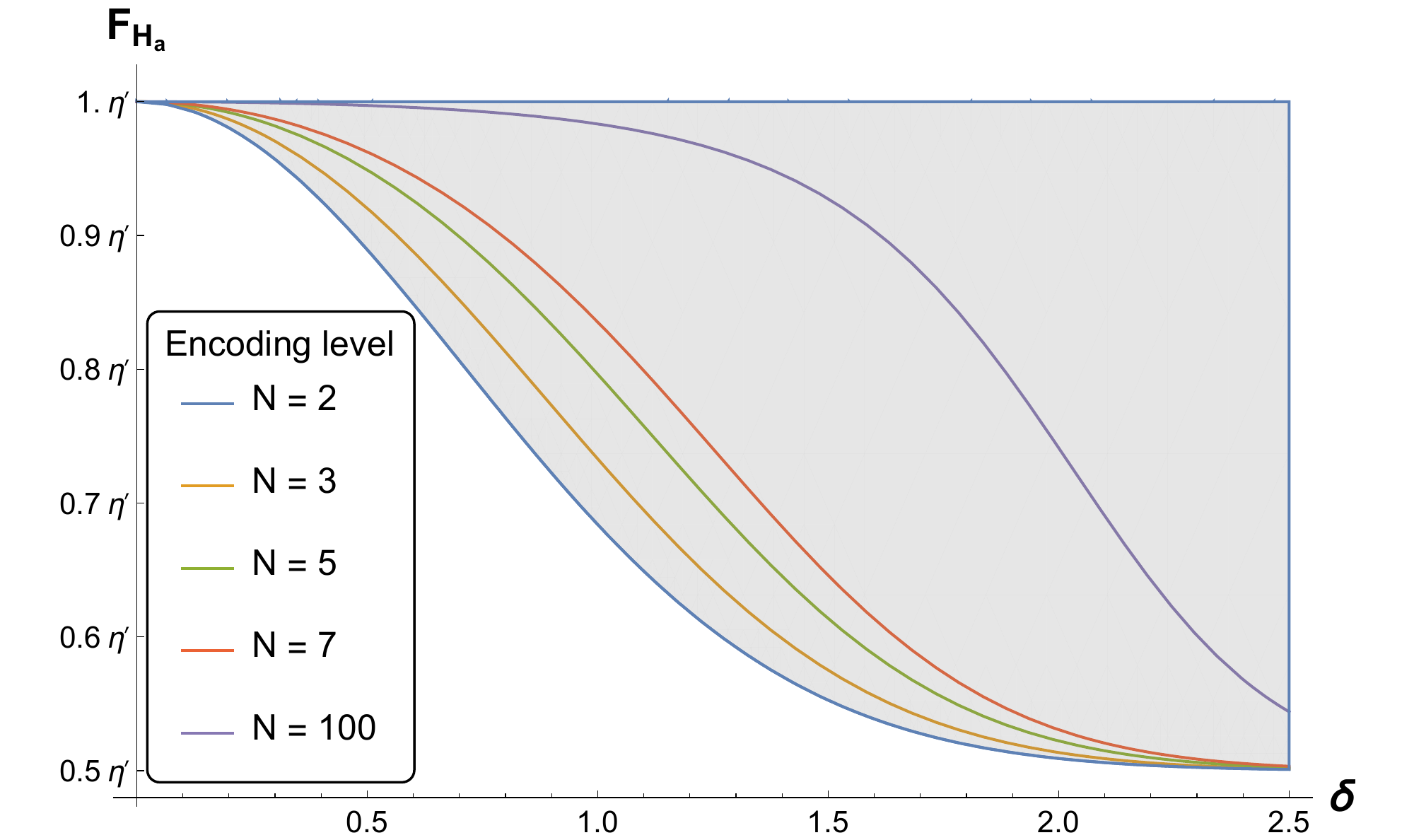} \\
    \includegraphics[width= \columnwidth]{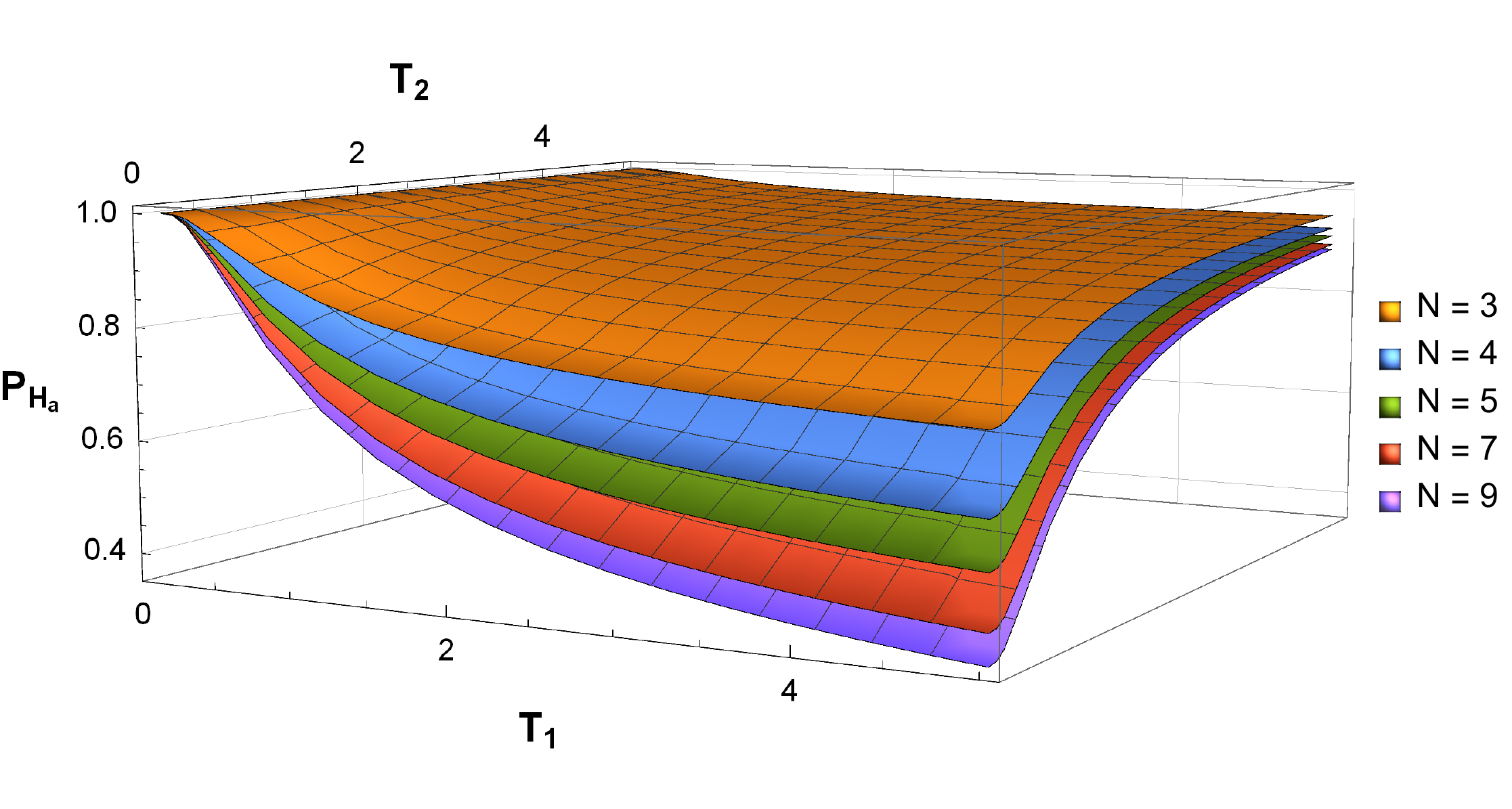}
    \includegraphics[width= \columnwidth]{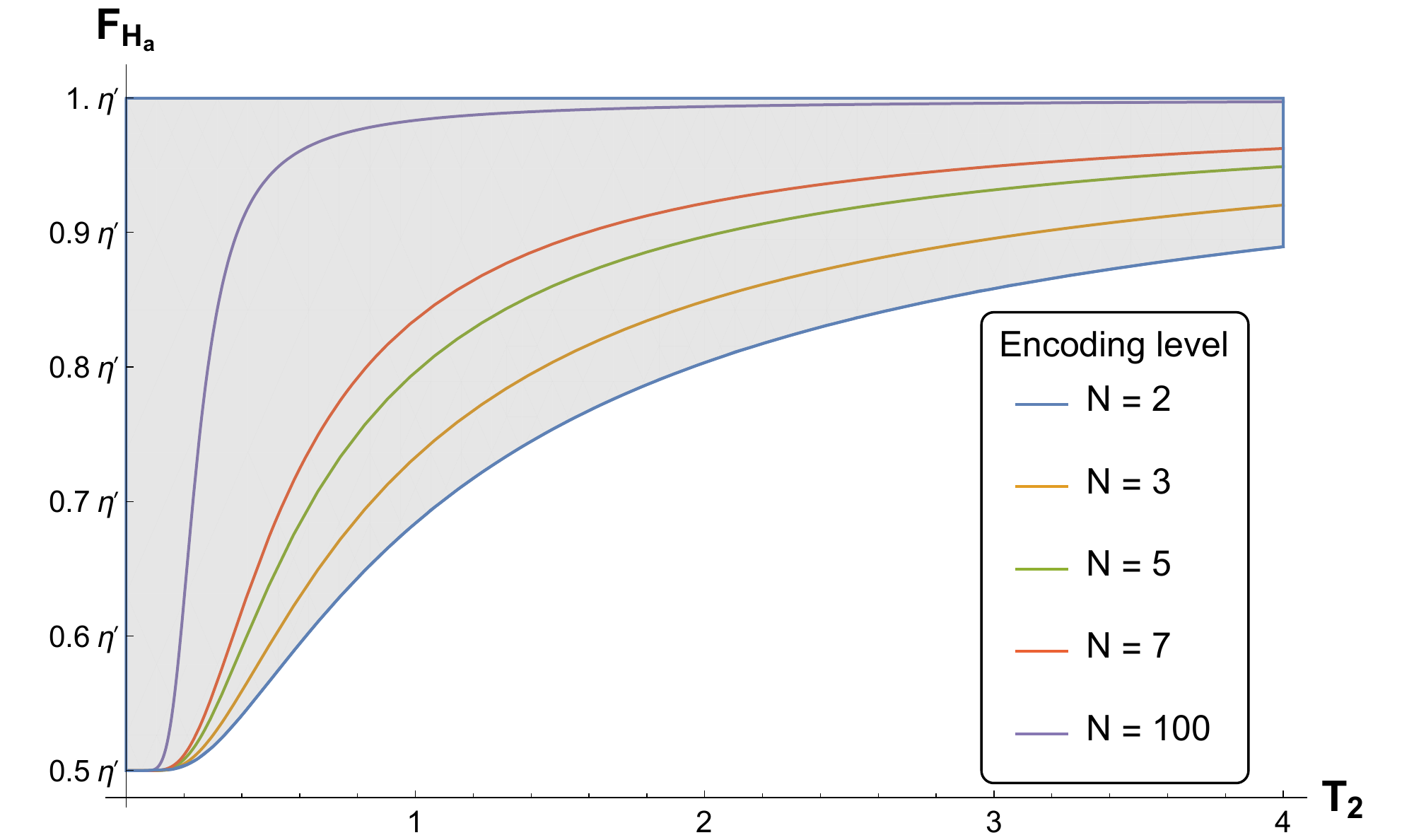}
    \caption{Analytic heralded error-correction results for the \textit{absence heralding} technique. Results for \textit{presence heralding} are given by simple transformations of these results (shift by $\eta$ for $P_H$, and scale by $(1-\eta)$ for $F_H$). (left) Heralding probability and (right) post-selected fidelity, parameterized in terms of loss-rate, and dephasing in terms of (top) phase-variance $\delta$, (bottom) $T_2$-time for $t_p =1$.}\label{fig:results}
\end{figure*}

The heralding probability and associated post-selected state fidelities are shown as a function of the channel parameters in Fig.~\ref{fig:results}, and the respective analytic expressions in Tab.~\ref{tab:analytic_summary}. We note that while we are specifically plotting for a Guassian noise model, the qualitative features can be expected to be the same for any i.i.d. error model. This is because depolarizing parameter $\lambda$ is related to the characteristic function $\phi_{p(\theta)}(z)$ through equation \eqref{eq:char_func}. A function and it's Fourier transform will have their variances inversely related like quadrature variances so even if the exact expressions for the fidelity and heralding probabilities might vary we can expect the qualitative behaviour to remain the same and the average map to be a depolarizing channel. 


\onecolumngrid

\twocolumngrid


\section{Single-qubit unitary operations} \label{sec:qubit_gates}

Once in the encoded basis, can we directly perform single-qubit unitary operations, without the rigmarole of decoding and encoding? The answer is yes. 

Consider the single qubit operation,
\begin{align}
    \ket{\psi_\mathrm{out}}_L = \hat{U} \ket{\psi_\mathrm{in}}_L,
\end{align}
in the logical qubit basis. In the encoded photonic basis, this can be expressed,
\begin{align}
    \ket{\psi_\mathrm{out}}_L &= \hat{Q}^\dag\hat{Q} [\hat{U} \oplus \mathbb{\hat{I}}_{N-2}] \hat{Q}^\dag\hat{Q} \ket{\psi_\mathrm{in}}_L,
\end{align}
where we have inserted the identity operation $\mathbb{\hat{I}}_{N-2}$ on the ancillary input photonic modes, and $\mathbb{\hat{I}}=\hat{Q}^\dag\hat{Q}$. This yields the equivalent redundantly-encoded photonic unitary operation (i.e between encoding and decoding),
\begin{align}
        \ket{\psi_\mathrm{out}}_\mathrm{enc} = \tilde{U} \ket{\psi_\mathrm{in}}_\mathrm{enc},
\end{align}
where,
\begin{align}
    \tilde{U} = \hat{Q} [\hat{U} \oplus \mathbb{\hat{I}}_{N-2}] \hat{Q}^\dag,
\end{align}
is the redundantly-encoded equivalent of the logical 2-qubit operation, obtained by conjugating with $\hat{Q}$.

\section{Discussion} \label{sec:discussion}

\subsection{Pros and cons}

Our scheme has the following advantages:

\begin{enumerate}
    \item It can be implemented in a number of quantum memory architectures such as atomic ensembles, optical cavities and delay lines.
    \item Any independent uncorrelated phase noise can be corrected for with sufficient levels of encoding. The physical source of these phase errors will depend on the particular architecture. For example, in a delay line temporal mismatch between photon arrival times will manifest as a phase error. If this is influenced by thermal fluctuations, it will be manifested as a dephasing error reflecting our theoretical calculations of the post error state. In an optical cavity array, the source of the phase error could be decay rate mismatch between cavities. All these cases are consistent with our model.
    \item Normally to correct phase mismatch one could either thermally or mechanically isolate the system or use a high intensity source to periodically measure phase errors and actively correct it using feedback. Our scheme mitigates this.
    \item Because we only need passive linear optics without feed-forward, this is quite scalable with present-day technology, notably integrated photonic waveguide chips.
    \item Robustness against mode loss. Standard QECs require the use of entangling gates such as CNOTs and the code state themselves can be highly entangled such as the GHZ states. These states however are not robust against loss in the sense of a partial trace operation while the W-state encoding will robust against such loss and increasingly so with higher levels of encoding. 
\end{enumerate}

The disadvantages of our scheme are:

\begin{enumerate}
    \item Inability to correct correlated phase fluctuations (e.g a uniform phase shift across all redundant memory cells).
    \item A multiplier in production cost and resource overhead, determined by the degree of redundancy.
\end{enumerate}

\subsection{Economic justification} \label{sec:economics}

There is a strong economic argument for the merit of our scheme, regardless of the state of engineering.

The main economic overhead associated with the protocol is the substitution of a single quantum memory with a bank of $N$ identical ones, a roughly linear cost overhead. However, via this trade-off, dephasing processes inherent within them can be asymptotically suppressed, enabling the construction of a quantum memory bank that overcomes the fidelity bounds of a single one.

Given that the engineering and production cost of a single quantum memory unit can increase exponentially with inverse infidelity
, an $N$-fold cost overhead is expected to be a more economically efficient mechanism for noise-suppression in the regime of very high fidelity targets.

The net cost of the memory bank scales roughly linearly with $N$, whereas the cost of a single cell within it grows exponentially with fidelity. The economically optimal configuration is determined by the minimum of this cost trade-off curve over $N$, for a given target fidelity $F_\mathrm{target}$,
\begin{align}
    C_\mathrm{net}(F_\mathrm{target}) \approx N\cdot C_\mathrm{unit}(F_\mathrm{unit}),
\end{align}
where $C_\mathrm{unit}(F_\mathrm{unit})$ is the engineering cost of a single memory cell with fidelity $F_\mathrm{unit}$, and the relationship between the target and unit fidelity follows from the respective heralded fidelity given in Tab.~\ref{tab:analytic_summary}, related by $N$. The crossover point, at which it becomes economically efficient to begin utilising our encoding, occurs when manufacturing a single memory cell with the target fidelity matches that of a redundant bank of cells with lower unit cost \cite{bib:DuanLukin01},
\begin{align}
    N\cdot C_\mathrm{unit}(F_\mathrm{unit}) \approx C_\mathrm{unit}(F_\mathrm{target}).
\end{align}

\subsection{Robustness against different noise models}

Whilst we have used a Gaussian noise model for detailed analysis in Section~\ref{sec:protocol}, this is by no means an absolute requirement on many of the results we present.  As mentioned in Section~\ref{sec:err_hrld} C, the property that determines the output state is the characteristic function for the random variables in the noise model for the unitary errors evaluated at $z=1$.  Due to the nature of the characteristic function, this value will be well defined in virtually all possible distributions, even ones that do not have a well defined moment generating function.  The exact details of specific properties of the scheme will change under different distributions (e.g. the $T_1$ and $T_2$ decoherence factors identified here won't be well defined in general), but the analysis from the point of view of the encoded state will be essentially the same as what we have presented here.

\subsection{Compatability with no-go theorem}

It might seem that this protocol permits the distillation of entanglement that is present within the input state using only operations that maintain the Gaussian form of a Gaussian state, which has been proven impossible by a no-go theorem \cite{Eisert2002}. The conditions for the no-go theorem do not apply here as the final heralding measurement, which heralds success on a (Gaussian) vacuum state, is overall not a Gaussian measurement under our proposal to measure in the Fock basis. The input Fock states are also non-Gaussian. This is similar to how the no-go Gaussian distillation theorem is avoided in current bosonic entanglement distillation schemes \cite{Browne2003, Ralph2009, Xiang2010, Walk2013a}.

\section{Comparison with other schemes}


The idea of error filtration in a passive linear optic network has been explored in \cite{bib:Gisin05_filt, li2007faithful,jiang2017self}. Broadly these schemes transmit a photon through a linear optical network such that some measurement outcomes will indicate an uncorrupted state in some output. We have formalized this intuition by giving an explicit code space and showed how it is robust against mode loss and i.i.d. dephasing noise. 

Other schemes have explored the use of probabilistic gates to protect against transmission loss such as \cite{ewert_ultra_2017} where optical Bell measurements are used along with a parity encoding. However the encoding states used for these schemes are highly entangled GHZ-like states. These states cannot be deterministically prepared using passive linear optics without introducing active feed-forward and additional photons to accommodate the higher level of encoding --- making such schemes highly impractical using present-day technology.

It is important to clarify the distinction between this protocol, which can be regarded as a form of error correction, and the more general concept of fault-tolerance where gate errors are accommodated for.

Here we have assumed that our encoding and decoding operations are ideal, and all the dephasing errors are associated with the channel between them. Furthermore, we are not considering full quantum computations, but rather the storage or communication of just a single photonic qubit.

While future work might consider the effects of errors in the encoding and decoding errors in this protocol, the presented analysis is nonetheless reasonably well justified in most practical circumstances.

Current linear optics technology, both using discrete elements or in integrated wave-guides, has become extremely mature and precise, enabling passive linear networks to be implemented with very high degrees of fidelity. 

On the other hand, photonic qubits communicated over long-distance links, via any medium, or which are held in quantum memories by coupling them to non-optical physical systems, are far more likely to contribute to these noise processes.

A further distinction between our scheme and conventional error correction schemes, is that we don't rely on any notion of code concatenation to asymptotically improve error thresholds. Instead, we directly expand our level of encoding by increasing the number of optical modes in the fan-out operation implemented by the QFT encoding operation.

Unlike most well-known codes whereby error syndrome measurements are used to apply feed-forward corrections to encoded qubits, this protocol does not rely on any form of active correction via syndrome extraction. Rather, dephasing noise is effectively mapped to non-determinism, such that upon success the effective dephasing rate has been reduced.

The final important distinction between this scheme and conventional QEC schemes, is that we do not create our encoded state via the introduction of additional qubits (i.e photons), but via the the introduction of additional optical modes, where the number of photons is preserved.

Owing to these conceptual differences compared to more familiar QEC and fault-tolerance techniques, our scheme as presented is especially suited to the context of photonic quantum communication or storage via coupling into quantum memories.

\section{Conclusion}

We have proposed a passive linear optics encoding, using W-states which have the property of being strongly robust against entanglement degradation from qubit loss. This encoding was shown to be robust against any dephasing error modelled as an uncorrelated independent and identically distributed dephasing process on each subsystem. We showed that the effective error probability is inversely related to the level of encoding $N$, vanishing in the large $N$ limit. The loss rate upper-bounds the fidelity and success probability, but its effect does not scale with $N$, given that uniform losses can be commute through passive linear optics systems.

The protocol is naturally suited to optical quantum memories (e.g via atomic ensembles, cavities, or delay lines), where the dominant error processes are independent dephasing and loss. Single-qubit operations are readily implementable within the encoded basis using conjugated, passive, linear optics operations.

We argued that for high-fidelity quantum memories, utilising this technique complimentary to improving engineering precision, has merit from an economic perspective, given the only linear overhead in cost associated with simple redundancy, versus the far greater cost of improving engineering precision.




\section*{Acknowledgements}

This work was supported by the Australian Research Council Centre of Excellence for Engineered Quantum Systems (Grant No. CE170100009), and Australian Research Council Centre of Excellence for Quantum Computation \& Communication Technology (Grant No. CE170100012). Peter Rohde is funded by an ARC Future Fellowship (project FT160100397). Austin Lund would also like to acknowledge fruitful discussions with Joshua Combes and Nicholas Menicucci in relation to this work. Madhav Krishnan Vijayan would like to thank Alexis Shaw for an insightful discussion on the physical source of errors.

\bibliography{bib}
\end{document}